\documentclass[a4paper,pre,10pt]{revtex4}
\usepackage[utf8]{inputenc}

\usepackage{amsmath,amsfonts,amssymb,bm}
\usepackage{graphicx}
\usepackage[hyperfigures,breaklinks,colorlinks,citecolor=blue,linkcolor=black]{hyperref}
\usepackage{mathbbol}
\usepackage[T1]{fontenc}
\usepackage{lmodern}
\usepackage[figurename=Fig.,labelfont=bf,labelsep=period]{caption}

\def\s{\sigma}

\def\p{\partial}
\def\bsigma{\bm{\sigma}}
\def\ud{\mathrm d}

\def\pij{p_{ij}}
\def\E{\mathrm{E}}

\begin{document}

\title{Dynamics of epidemics from cavity master equations}
\author{Ernesto Ortega}
\author{David Machado}
\author{Alejandro Lage-Castellanos}
\email{ale.lage@gmail.com}
\affiliation{Complex Systems Group, Physics Faculty, Havana University, Cuba}

\begin{abstract}
We apply the Cavity Master Equation to epidemic models, and compare it to previously known approaches. We show that CME corrects some equations in dynamic message passing and also outperforms this approach as well as a couple of mean field approaches. We explore average case predictions and extend the cavity master equation to SIR and SIRS models, applying the CME for the first time to models with more than 2 states.
\end{abstract}

\maketitle

\section{Introduction}

Since the seminal works introducing susceptible-infectious-recovered compartment models (SIR) of Kermack and McKendric \cite{kermack1927contribution}, epidemics modeling has grown fast as a field. The approach has changed with time, from dynamical systems or population dynamics towards a more stratified approaches as patchy, mobility based, age-structured or contact matrices compartment models.

The current context of a global COVID-19 pandemic and the perspective of a coexistence with an endemic virus requires a test-trace-isolate epidemiological system to keep the outbreak controlled. Much attention is now put on agent based models \cite{kerr2020covasim, cuevas2020agent, hinch2020openabm} that could improve the efficacy of the testing strategy. Assuming that new technologies can provide reliable contact data between humans, the likelihood of people being infected needs to be estimated either by numerical simulations or some statistical modeling. To this end, it is suitable to count with fast algorithms that can accurately predict probabilities of infection for agents in networks.

There are a wide variety of such algorithms. The classical  approach to the forecasting of epidemics on networks is an averaging of the master equation of the process complemented by a factorization assumption at some level. This yields a hierarchy of ever more complex but more accurate differential equations for expected values and correlations\cite{simon2011exact, sahneh2013generalized}. Most of the time only the first two levels, known as individual based mean field and pair based mean field, are used.

More recently, ideas from discrete optimization algorithms have sneaked into the inference of SIR kind of models in the shape of Dynamical Message Passing \cite{lokhov2014inferring, shrestha2015message} or Belief Propagation (BP) \cite{altarelli2014bayesian, braunstein2016inference}. The main difference with respect to the previous approach is the appearance of conditional -rather than multivariate- probabilities to be integrated in time.
It has been used with success in the reconstruction of epidemics on graphs and it is currently being tried in the task of risk assessment for COVID-19 \cite{baker2020epidemic,messageGinestra}. However, these two types of approaches, the standard master equations and the message passing one has remained similar but theoretically disconnected.

It is known that BP fixed point is connected to the cavity method from statistical mechanics \cite{krauth1989cavity}. An extension of cavity method to continuous time Markov Chain processes within discrete spin systems have been recently achieved through the derivation of a set of differential equations for cavity conditional probabilities: the cavity master equation (CME) \cite{aurell2017cavity}. In this article we explore CME's application to SIS and SIR-like models in graphs (section \ref{sec:CMESIS}). We will show that CME somehow binds both approaches, since it starts from a master equation, but produces message passing-like equations. CME seems to be the formal path to obtaining (and fixing) the dynamic message passing of \cite{shrestha2015message} that were intuitively presented then.

When compared with Monte Carlo simulations of the epidemic, CME is shown to outperform not only rDMP but also a naive individual based and a kind of pair based mean field approaches. In some simple cases we draw analytical results for a group of steady state quantities and the corresponding critical spreading rate (section \ref{sec:steady}). We also explore the average case dynamics of these equations, in particular in random regular and Erdos-Renyi graphs (section \ref{sec:average}). We conclude by presenting an extension of the Cavity Master Equation to the case of sequential multi-state models and how this applies to to SIR and SIRS models in section \ref{sec:SIRSIRS}. 




\section{Epidemics on networks} \label{sec:epidemics}

In what follows we focus on continuous time compartment epidemic models on networks. We assume a fixed network of contacts to be given $G=(V,E)$ with a set of vertices $V=\{1,2\ldots,N\}$ and a set of edges $E$. An edge $(i,j)$ is present if nodes $i$ and $j$ are neighbors in the network, meaning there is a possibility of transmission of diseases between both nodes.
\begin{figure}
 \includegraphics[width=0.3\textwidth, angle=0]{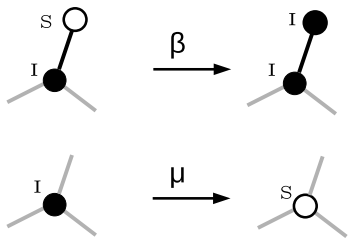} \caption{Allowed transitions in SIS compartment model on networks. \label{fig:SISdiagram}}
\end{figure}

The standard susceptible-infectious-susceptible model (SIS) considers the nodes to be in either of two compartments (states) $X_i = 0 \equiv \mbox{susceptible}$, or $X_i = 1 \equiv \mbox{infectious}$ and is the simplest standard for recurrent transmissible diseases. The epidemic is thus a continuous time stochastic process with only two admitted transitions occurring at
\begin{itemize}
 \item rate $\beta$, at which a link $(i,j)$ can transmit the state $1$ from node $i$ to $j$;
 \item rate $\mu$ at which state $1$ decays to state $0$ on any infectious node,
\end{itemize}
as represented in diagram \ref{fig:SISdiagram}. An analytical description of this stochastic process is given by the master equation for the evolution in time of the probability over the whole configuration space $P(X_i,\ldots,X_N\:,t)$ \cite{sahneh2013generalized}. However concise and exact, the integration in time of such equation is generally impractical given the size $2^N$ of the configuration space. 

Attempts to reduce the complexity start from factorizing the single master equation into many equations for each node marginals $P_i(X_i,t)$. Given that the $X$'s are two-state variables, $P_i(X_i)$ is parameterized by the mean value $\E [X_i] = P_i(X_i=1)$. This results in an equation that is still exact \cite{sahneh2013generalized}
\begin{equation}
\frac{\ud \E\left[  X_{i}\left(  t\right)  \right]  }{\ud t}  =\E\left[  -\mu
X_{i}\left(  t\right)  +\left(  1-X_{i}\left(  t\right)  \right)  \beta
\sum_{j=1}^{N}a_{ij}X_{j}\left(  t\right)  \right]
\label{eq:meanXdt}
\end{equation}
where $a_{ij}$ are the elements of the adjacency matrix, meaning that $a_{ij} = 1$ if nodes $i$ and $j$ are neighbors ($(i,j)\in E$) and is zero otherwise.

However, the expectation value on the right hand side acts over products of variables $X_i(t) X_j(t)$, which requires a differential equation for the evolution of the correlations. Not surprisingly, the two point correlations functions depend on three point correlations and so on and so forth. 

The simplest closure of equation eq.  (\ref{eq:meanXdt}) is the individual-based mean field (IBMF) in which independence is assumed $\E[X_i(t) X_j(t)] \approx \E[ X_i(t)] \: \E [ X_j(t)] \equiv \rho_i(t) \rho_j(t)$ and therefore eqs. (\ref{eq:meanXdt}) are now a closed set of non linear differential equations:
\begin{equation}
\frac{d \rho_i(t)  }{dt}=-
 \mu \, \rho_i(t)  +\beta [1-\rho_i(t)] \sum_{j \in \partial i} 
\rho_j(t)
\label{eq:IBMFSISequations}
\end{equation}
where we used the notation $\partial i$ to represent the set of neighbors of node $i$.

The second simplest closure is the one known as pair-based mean field (PBMF), in which two point correlations are treated analytically \cite{cator2012second, mata2013pair}.
\begin{eqnarray}
\frac{\ud \E\left[  X_{i}X_{j}\right]  }{dt}  
&  =-2\mu \E[X_{i}X_{j}]+\beta\sum_{k=1}^{N}a_{ik}\E[X_{j}X_{k}]+\beta
\sum_{k=1}^{N}a_{jk}\E[X_{i}X_{k}]-\beta\sum_{k=1}^{N}(a_{ik}+a_{jk})\E[X_{i}X_{j}X_{k}]
\label{eq:EXiXidt}%
\end{eqnarray}
but a factorization is assumed for higher correlations. Different approaches have been used to approximate $\E[X_i X_j X_k]$ in terms of smaller correlations. In this paper we will compare with the approximations proposed in \cite{cator2012second} $\E[X_i X_j X_k] \approx \E[X_i X_j]\E[X_k] \equiv \phi_{ij}(t) \, \rho_i(t)$:
\begin{eqnarray}
 \frac{\ud\rho_{i}(t)}{\ud t} &=& -\mu \rho_{i}(t) + \beta \sum_{j \in \partial i} \phi_{ij}(t)  \label{eq:rhoi_dt_PBMF} \\
\frac{\ud \phi_{ij}(t)}{\ud t} &=& -(2\mu+\beta) \phi_{ij}(t) + \mu \rho_i(t)-  \beta \phi_{ij}(t) \sum_{k \in \partial i \setminus j} \rho_{k}(t)  \nonumber \\ & & + 
\beta \left[ 1 - \rho_{i}(t) - \phi_{ij}(t) \right] \sum_{k \in \partial j \setminus i} \rho_{k}(t) \label{eq:joint_PBMF1}
\end{eqnarray}
and \cite{mata2013pair} $\E[X_i X_j X_k] \approx \frac{\E[X_i X_j]\E[X_j,X_k]}{E[X_j]} \equiv \frac{\phi_{ij}(t) \, \phi_{jk}(t)}{\rho_{j}(t)}$.
\begin{eqnarray}
 \frac{\ud\rho_{i}(t)}{\ud t} &=& -\mu \rho_{i}(t) + \beta \sum_{j \in \partial i} \phi_{ij}(t)  \nonumber \\
\frac{\ud \phi_{ij}(t)}{\ud t} &=& -(2\mu+\beta) \phi_{ij}(t) + \mu \rho_i(t)  -  \beta \phi_{ij}(t)/(1-\rho_{i}) \sum_{k \in \partial i \setminus j} \phi_{ik}(t)  \nonumber \\ & & + 
\beta \left[ 1 - \rho_{i}(t) - \phi_{ij}(t) \right] \sum_{k \in \partial j \setminus i} \phi_{jk}/(1-\rho_{j})(t)
\label{eq:joint_PBMF2}
\end{eqnarray}

We will refer to the two different approximations shown in (\ref{eq:joint_PBMF1}) and (\ref{eq:joint_PBMF2}) as {\bf PBMF-1} and {\bf PBMF-2}, respectively. 

In both approaches, IBMF and PBMF, the expected values evolving in time are intended to be expectations over different stochastic stories of the whole epidemic process. Therefore they are to be compared with averages over many Monte Carlo simulations of such process.

A slightly different approach to modeling epidemics on graphs come from message-passing inspired methods. The dynamic message passing \cite{lokhov2014inferring, shrestha2015message} involves a set of probabilities and a set of conditional probabilities, rather than correlations \cite{shrestha2015message}:
\begin{eqnarray}
  \frac{\ud p_i }{\ud t} &=&  - \mu p_i + \beta (1-p_i) \sum_{k} p_{ki} \label{eq:moorepi}\\
    \frac{\ud \pij}{\ud t} &=&  - \mu \pij \quad + \quad (1-p_j) \beta \sum_{k \in \p i \setminus j}  p_{ki} \label{eq:moorepij}
\end{eqnarray}
where $p_i = \E [X_i]$ similarly as before, but $p_{ij} \equiv P(X_i = 1 | X_j = 0)$ is a conditional probability resembling the kind of cavity fields or messages that commonly appear in the cavity method and the belief propagation algorithm for discrete optimization. 

The main results of this article are drawn from the cavity master equation \cite{aurell2017cavity}. They are quite similar to those of rDMP and help in formalizing this rather empirical approach by deriving it from a more solid mathematical setting. In doing so we do not only correct one term of the rDMP equations, but also underline the approximations involved and therefore shed light on possible improvements.

\section{Cavity Master equations for SIS epidemics}
\label{sec:CMESIS}

In order to connect with its first presentation in \cite{aurell2017cavity}, we start by considering the general continuous time dynamics of a system $\bsigma = \{\s_1,\ldots,\s_N\}$ of $N$ bimodal variables $\s_i \in\{\pm 1\}$ interacting with their neighbors in some given topology. In the very generic Markovian case, the dynamic is fully defined by the rate function $r_i(\bsigma)$ at which variables flip their states from $\s_i \to -\s_i$. The distribution $P(\bsigma,t)$ in the configuration space of this stochastic process is ruled by the joint master equation:
\begin{equation}
\frac{d P (\bsigma) }{dt} = - \sum_{i=1}^N \Big[
r_i(\bsigma) P (\bsigma) - r_i(F_i(\bsigma)) P (F_i(\bsigma) ) \Big]\, ,
\label{eq:originalME}
\end{equation}
where $F_i$ represents the flip operator on variable $i$, \emph{i.e.} $F_i (\bsigma) = \{\sigma_1, \dots, \sigma_{i-1}, -\sigma_{i},\sigma_{i+1}, \dots, \sigma_N\}$.

Although exact, the previous equation is useless already for middle size systems, since it actually represents a set of $2^N$ coupled differential equations that take exponential time to enumerate, let alone to integrate. However this is the correct starting point for approximations, as is usually done to obtain  mean field equations (\ref{eq:IBMFSISequations},\ref{eq:rhoi_dt_PBMF},\ref{eq:joint_PBMF1},\ref{eq:joint_PBMF2}) in epidemics models \cite{sahneh2013generalized, simon2011exact, sharkey2008deterministic}.

In \cite{aurell2017cavity} this master equation is recast into an equilibrium problem by extending the configuration space to consider the continuous trajectory of each variable in time ${\bf X} = \{X_1,\ldots,X_N\}$ where $X_i = \{\s_i(t):\forall_{t\in[o,T]}\}$. Although at glance it seems untreatable the infinite dimensional space for the functions $X_i(t)$, the discrete values of $\sigma_i(t)$ allows for a codification of the functions in a numerable number of transitions times $\{T_1^{\{i\}},T_2^{\{i\}},\ldots\}$ such that $ \sigma_i(T_k^{\{i\}}) = - \sigma_i(T_k^{\{i\}}+\ud t)$.  The resulting Random Point Process is treated with standard techniques in equilibrium statistical mechanics to write down a closed set of cavity master equations as:
\begin{eqnarray}
\frac{\ud P (\sigma_i) }{\ud t} &=& - \sum_{\sigma_{\p i}} \Big[
r_i(\sigma_i, \sigma_{\p i}) \big[ \prod_{k \in \p i }
p(\sigma_k| \sigma_i) \big] P(\sigma_i)
-  r_i(-\sigma_i, \sigma_{\p i}) 
\big[ \prod_{k \in \p i }
p(\sigma_k| -\sigma_i) \big]P(-\sigma_i) \Big]
\label{eq:CMEPi} \\
 \frac{\ud p(\s_i|\s_j)}{\ud t} &=& -  \sum_{\sigma_{\partial i\setminus j}} \Bigg[ r_{i}[\sigma_i,\sigma_{\partial i}]  \Big[\prod_{k\in\partial i \setminus j } p(\sigma_k|\s_i)\Big] p(\s_i|\s_j) 
-  r_{i}[-\sigma_i,\sigma_{\partial i}] \Big[\prod_{k\in\partial i \setminus j } p(\sigma_k|-\s_i)\Big] p(-\s_i|\s_j) \Bigg] \label{eq:CMEpij}
\end{eqnarray}
We have lighten the notation by not putting the $i$ and $ij$ dependence of the distributions, understanding that they assume the index of the variables they depend on (as $P(\sigma_i)\equiv P_i(\sigma_i,t)$ and $p(\sigma_i|\sigma_j) \equiv p_{i,j}(t,\sigma_i|\sigma_j) $). Both probabilities are intended in the sense ``over the ensemble of dynamic evolutions up to time $t$'', starting from the same initial conditions.

This is a substantial improvement over the original master equation since we are dealing now with $N$ functions $P_i(\sigma_i,t)$ representing the distribution of variables $\s_i$, and with $N*\langle k \rangle$ functions $p_{(i,j)}(\s_i|\s_j,t)$ that represent the probability of finding variables in a given state, conditioned to the state of one of its neighbors ($\langle k \rangle$ is the average degree of the nodes in the interactions network). In the worst case of a fully connected system, we still would have $O(N^2)$ equations, that can be numerically integrated even for relatively large systems, compared to what we can do with eq. (\ref{eq:originalME}).



We can readily translate the cavity master equations (\ref{eq:CMEPi},\ref{eq:CMEpij}) to the case of S.I.S epidemic model by identifying our two states as $S\to \sigma_i= -1$ and $I \to\sigma_i = 1$. 
After complementarity ($P_i(I)+ P_i(S) =1$), it is enough to track the infection probability in each node $P_i(I)$. The first term in equation (\ref{eq:CMEPi}) largely simplifies due to the fact that the recovery rate $r_i(\sigma_i=1, \sigma_{\p i}) = \mu$ is independent of the neighbors state, resulting in $- \mu P(I)$. Considering also that the transmission rates are additive $r_i(S,\sigma_{\p i}) = \sum_{k'} r_i(S,\sigma_{k'})$ and that $r_i(S,S) = 0$ and $r_i(S,I) = \beta$ we get to the cavity master equations for the S.I.S model as
\begin{eqnarray}
  \frac{\ud p_i }{\ud t} &=&  - \mu p_i + \beta (1-p_i) \sum_{k} p_{ki} \label{eq:CMESISpi}\\
    \frac{\ud \pij}{\ud t} &=&  - \mu \pij \quad + \quad (1-\pij) \beta \sum_{k \in \p i \setminus j}  p_{ki} \label{eq:CMESISpij}
\end{eqnarray}
where we simplified notation further by making $P(\s_i=1) \equiv p_i$ and $P(\s_i=1 | \s_j=-1) \equiv \pij$. Appendix \ref{ap:average} contains a more detailed derivation of these equations for SIS model.

\begin{figure}
 \includegraphics[width=0.43\textwidth]{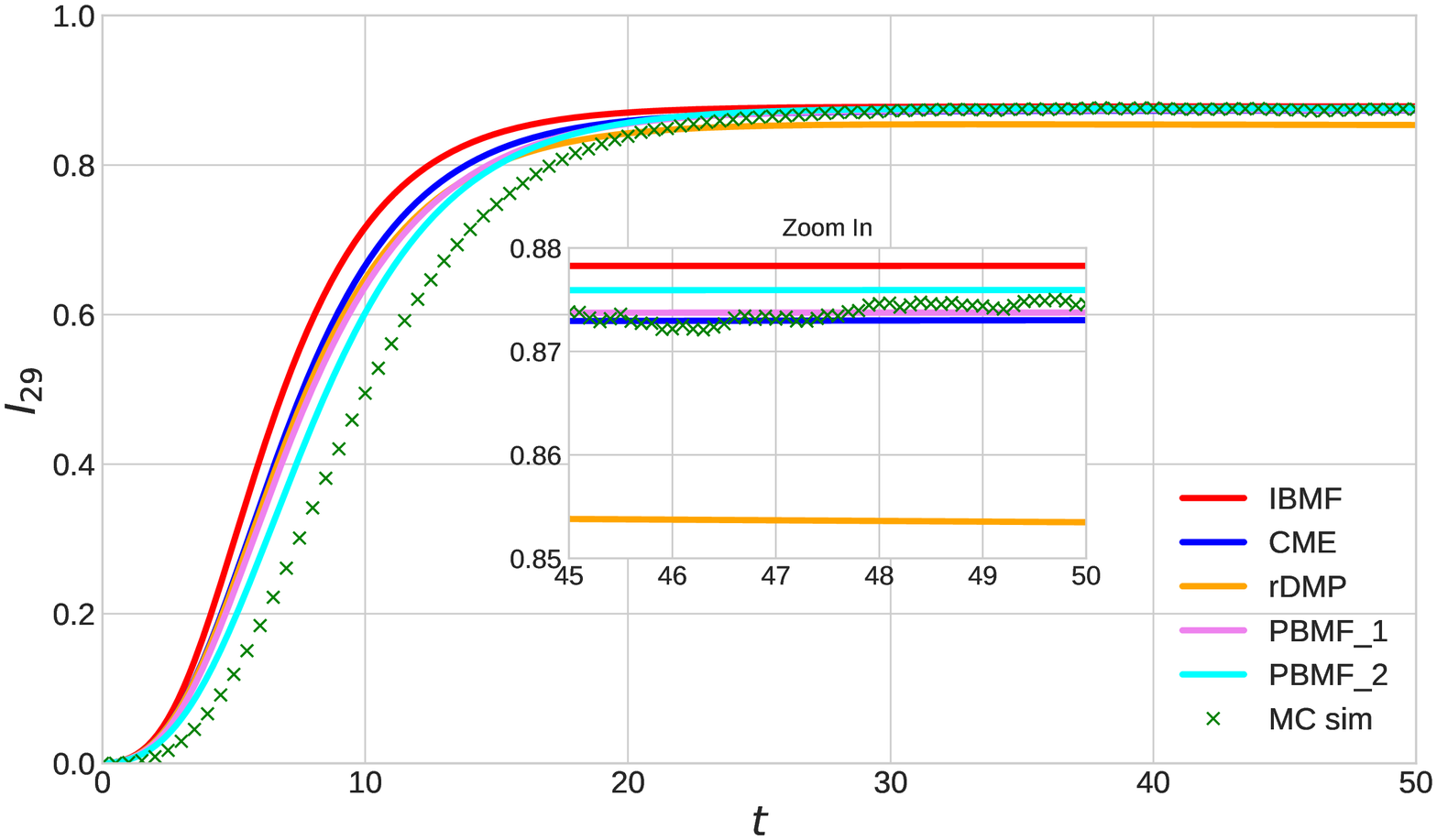}
 \includegraphics[width=0.43\textwidth]{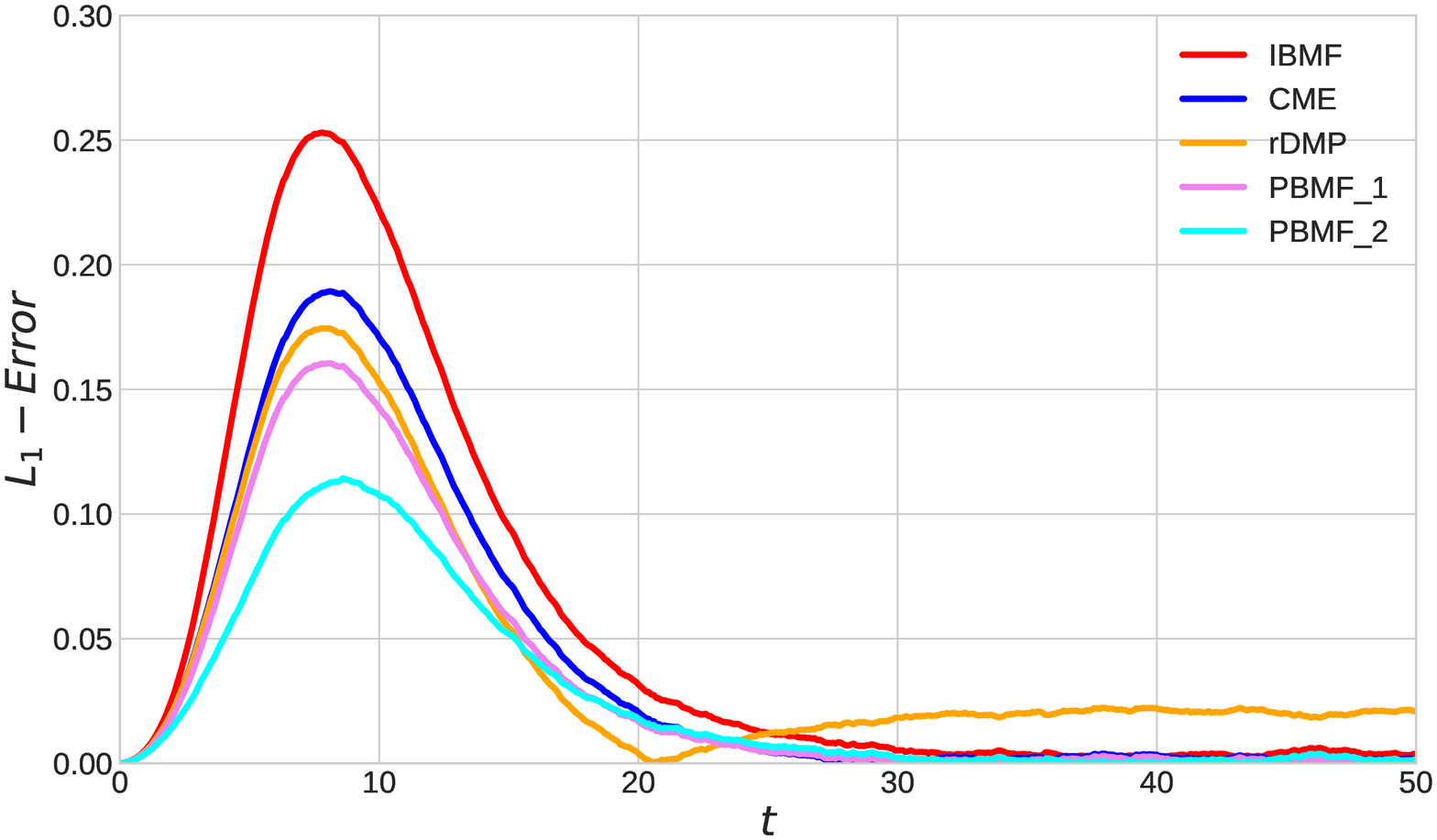}
 \caption{ \textbf{Left:} Probability of node 29 to be infected as a function of time, discounting at each time the situations in which the epidemics disappear. The epidemic outbreak was in node 1, with $\beta=0.1$ and $\mu=0.05$. The inset is a closer look in the stationary state. \textbf{Right:} L1 distance between the four approximations IBMF, PBMF, rDMP, CME with respect to $10^5$ MC simulations.  \label{fig:zacSIS}}
\end{figure}

Equations (\ref{eq:CMESISpi}) and (\ref{eq:CMESISpij}) are almost identical to rDMP (\ref{eq:moorepi}) and (\ref{eq:moorepij}), except for the very last term where $(1-p_j)$ is replaced by $(1-p_{ij})$. Not surprisingly, the results from both approximations might not be too different in certain cases. For instance, an SIS epidemic outbreak in the Zacharia's karate club network  (following  \cite{shrestha2015message}), starting at node 1 both approaches are quite similar as shown in figure \ref{fig:zacSIS} (left). In figure \ref{fig:zacSIS} (right) the L1 distance between the average from many Monte Carlo simulations and the predictions made by all four methods CME, rDMP, IBMF and PBMF shows features that will repeat in other benchmarks:
\begin{itemize}
 \item that both rDMP and CME get the general qualitative behavior well, 
 \item with a rather faster outbreak expansion in the transient, compared to Monte Carlo simulations, and
 \item with CME better fitting the stationary state, and quite close (exactly the same for regular graphs) to PBMF-2. 
\end{itemize}
Individual based mean field tends to be the fastest growing prediction. Overall, PBMF-2 seems to be the most accurate approximation, both for the transient and the steady.

When computing the Monte Carlo averages we have neglected the simulations in which the epidemic is randomly wiped out in the first few iterations. None of these methods can take this fluctuations into account, since they are all mean-field approaches.

\begin{figure}
 \includegraphics[width=0.44\textwidth]{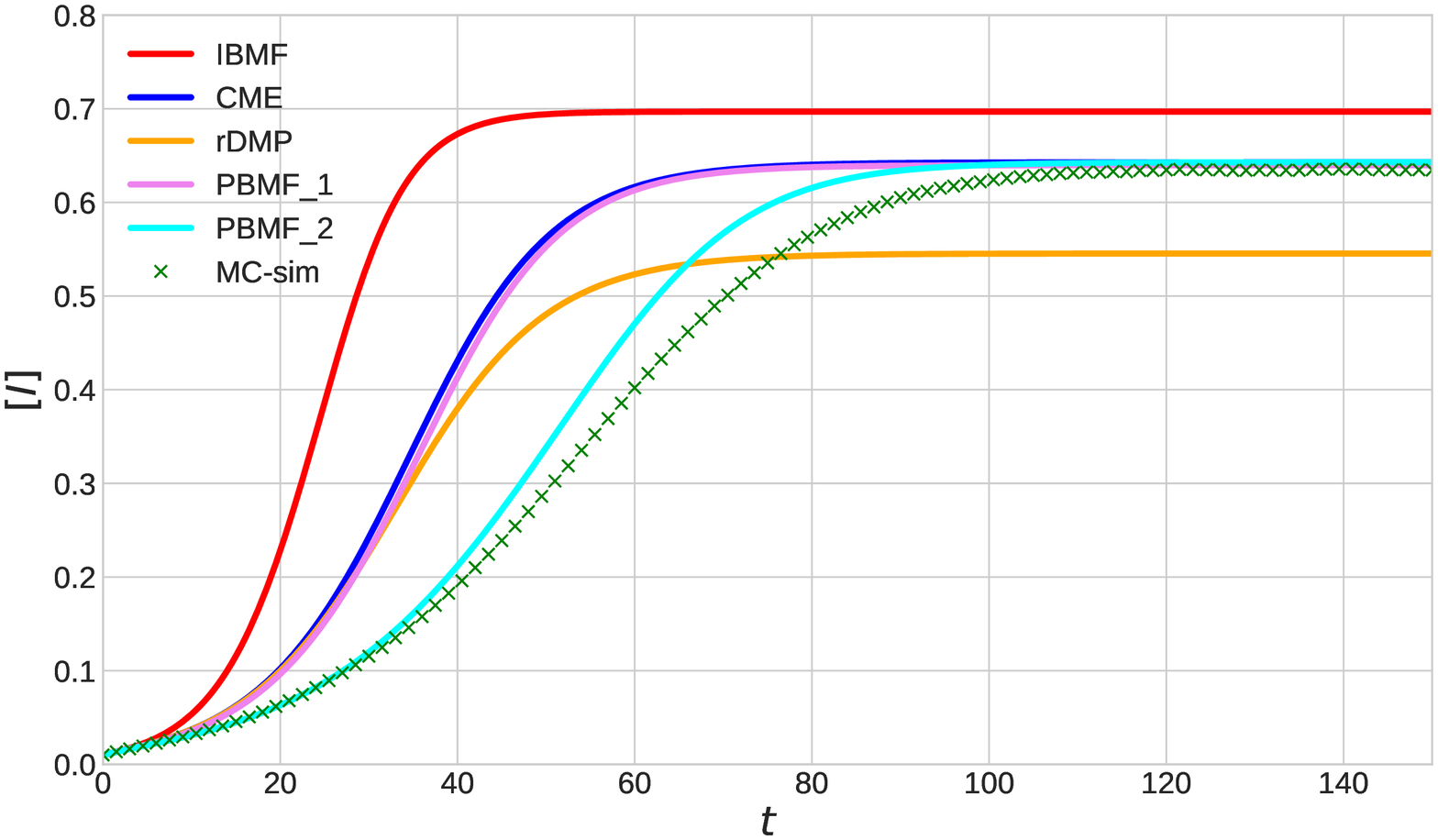}\includegraphics[width=0.44\textwidth]{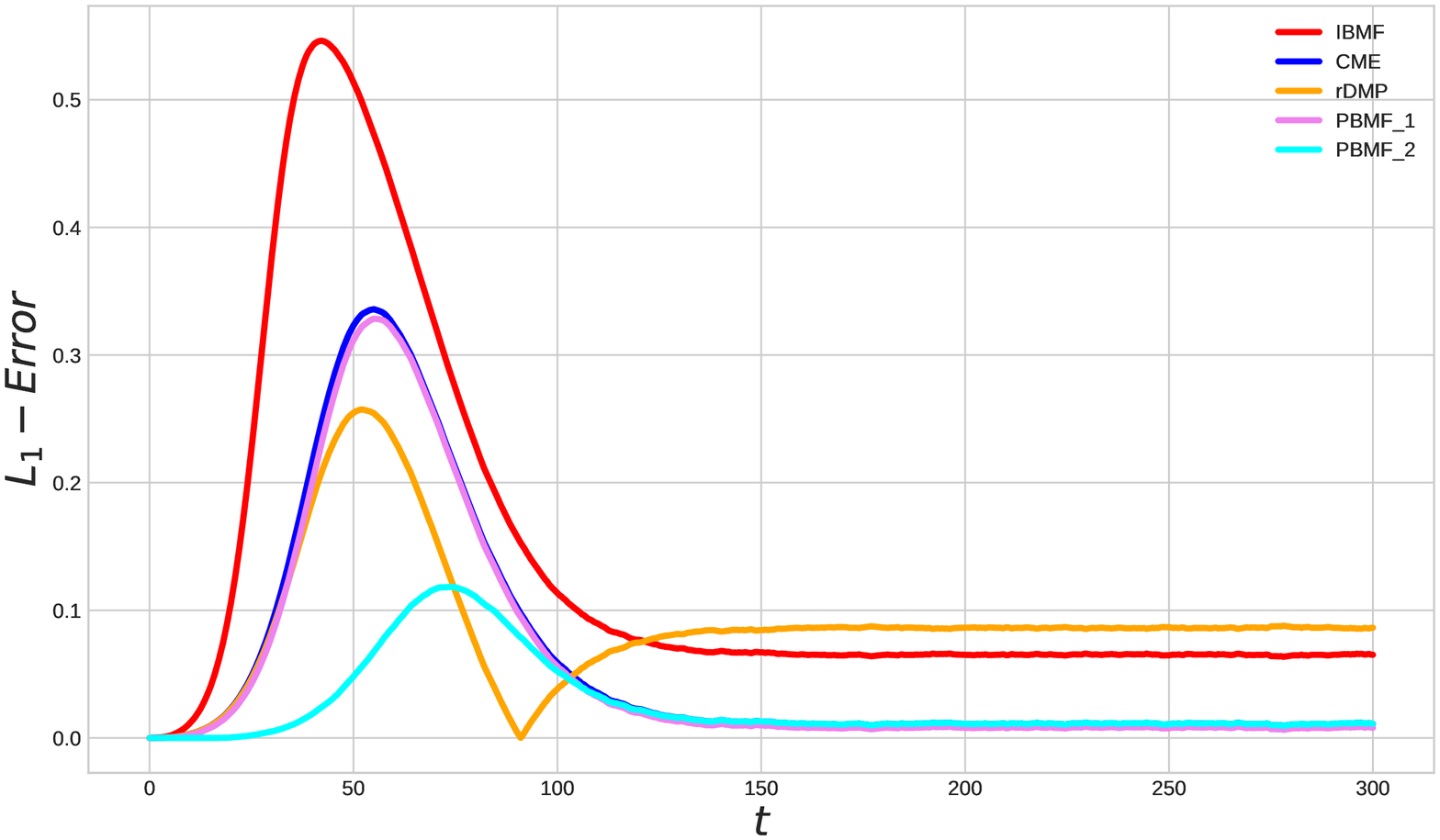}
\includegraphics[width=0.9\textwidth]{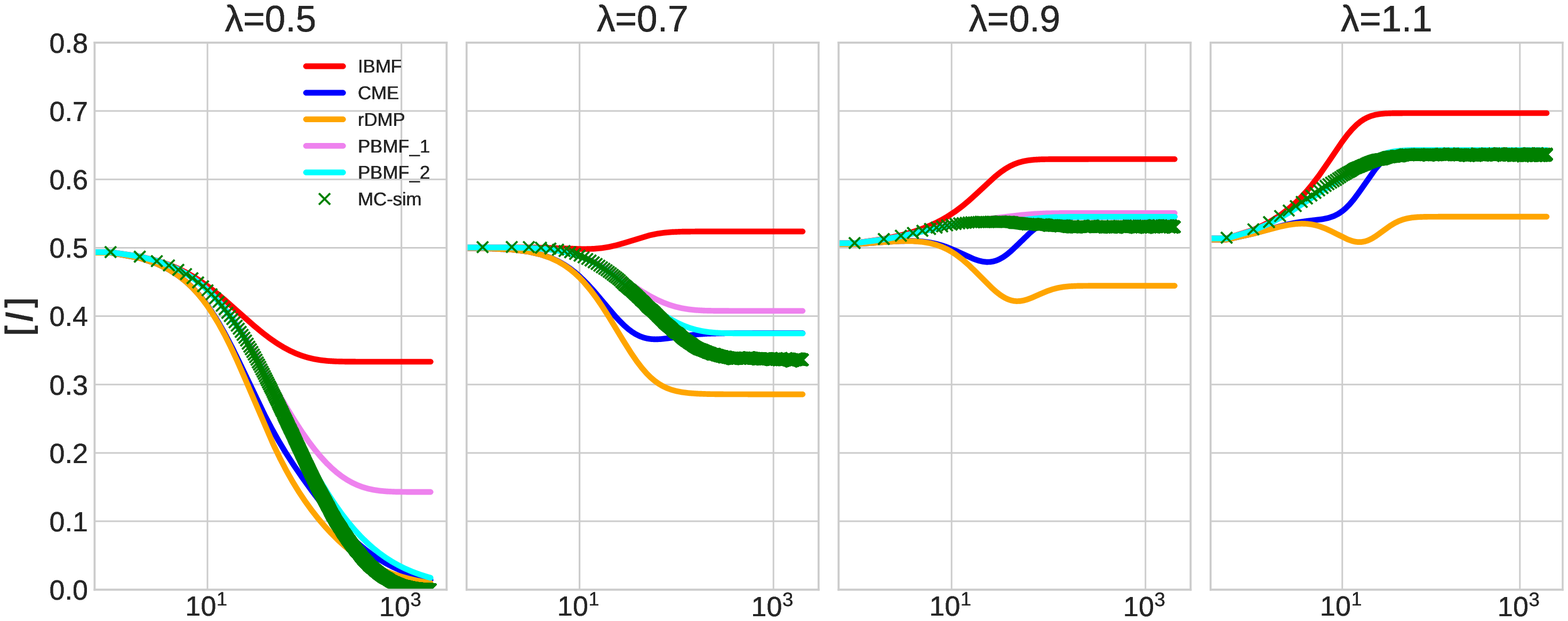}
 \caption{ {\bf Random regular graph} with connectivity $k=3$: {\it Top figures}- Percent of the population in the infected state (left) and the $L_1$ distance (right), in a graph of 1000 nodes with parameters $\beta= 0.11$ and $\mu = 0.1$ and an outbreak in one node of the system. {\it Bottom figures}: Percent of the population in the infected state for different values $\lambda=\beta/\mu$ in a graph of 1000 nodes, starting with half of the population in the infected state. In all figures each point of MC is an average over $10^4$ simulations}
 \label{fig:rand_reg_k3}
\end{figure}

Differences are more evident in the case of random regular graphs with degree $k=3$ in figure \ref{fig:rand_reg_k3}. We represent the average epidemic size, i.e. the fraction of infected nodes with respect to $N=1000$ the number  of nodes in the graph at every time step. We started from a  fraction $\alpha = 0.5$ of nodes infected, and observed the onset of the endemic state.

\section{Endemic (steady) state }
\label{sec:steady}
We analytically compute the epidemic size in the endemic state for these approximations for random regular graphs. Since every node has the same degree $k$, the equations are similar for every node, and we can assume that in the steady state the topology is averaged out, and all the probabilities are the same, regardless the node indexes.

Working explicitly for the CME approximation, stationarity means we have to set $\frac{\ud p_i }{\ud t}$ and $\frac{\ud \pij }{\ud t}$ to $0$ in equations (\ref{eq:CMESISpi}) and (\ref{eq:CMESISpij}):
\begin{eqnarray}
  \mu p_i &=&  \beta (1-p_i)\sum_{k} p_{ki} \equiv \beta (1-p_i) k \pij \qquad  \mu \pij = (1-\pij) \beta \sum_{k \in \p i \setminus j}  p_{ki} \equiv (1-\pij) \beta (k-1)
\end{eqnarray}
%
where now the indices $i$ and $i,j$ are generic. Solving this system of equations we get for the two variables $p_i$ and $p_{ij}$:
\begin{eqnarray}
 p_i &=&\frac{\lambda(k-1)-1}{\lambda(k-1)-1+(k-1)/k} 
  \qquad   \pij =  1- \frac{\mu}{(k-1)\beta}  
\end{eqnarray}
This can be already tested against simulations of random regular graphs. Furthermore we can obtain the spreading rate or effective infection rate $\lambda = \beta/\mu$ above which there is an endemic outbreak (sustained in time epidemic) by solving $p_i(k) = 0$, resulting in the epidemic threshold $\lambda_c = 1/(k-1)$. A similar procedure for all five approximations results in table \ref{tb:table_endemic_state}. Notably, in spite of being different in the transient, the pair-based mean-field 2 approach coincides in the stationary state with the cavity master equation. 

The critical values $\lambda_c=1/k$ for IBMF are consistent with those in  \cite{pastor2015epidemic} and citations therein with similar approaches. PBMF-2, CME and rDMP prediction $\lambda_c = 1/(k-1)$, however, is known to be a second order correction to the endemic threshold \cite{mata2013pair,cator2012second}, and it is numerically \cite{mata2013pair,ferreira2012numeric} seen to outdo the individual based value.

\begin{center}
\begin{table}
\def\arraystretch{2.5}
\begin{tabular}{|c|c|c|}
 \hline Approximation & Endemic state (equilibrium) & $\lambda_c$ \\
\hline
 IBMF & \(p_i =  1 - \frac{1}{k\lambda} \) & $\lambda_c = \frac{1}{k}$\\
\hline
 PBMF-1 & $p_i = \frac{k+k(k-1)\lambda-2/\lambda -1}{(k-1)(\lambda k +2)}$ & $ \lambda_c = \frac{\sqrt{1+\frac{8k}{k-1}}-1}{2 k} $\\
\hline
 PBMF-2 & $p_i = \frac{\lambda(k-1)-1}{\lambda(k-1)-1+(k-1)/k}$ & $ \lambda_c = \frac{1}{k-1} $\\
\hline
 rDMP & $\begin{array}{lcl}  
 p_i &=&  1- \frac{1}{(k-1)\lambda} \\
  \pij &=&   \frac{(k-1) \lambda-1}{k \lambda} 
\end{array}$
 &  $\lambda_c = \frac 1 {k-1}$\\
\hline
 CME & $\begin{array}{lcl}
p_i &=&\frac{\lambda(k-1)-1}{\lambda(k-1)-1+(k-1)/k}  \\
  \pij &=&  1- \frac{1}{(k-1)\lambda}  
\end{array}$
 & $\lambda_c = \frac 1 {k-1}$ \\ 
 \hline
\end{tabular}
 \caption{Steady state and epidemic threshold $\lambda_c$ under four approximations: IBMF, PBMF, rDMP and CME}
 \label{tb:table_endemic_state}
 \end{table}
\end{center}

\begin{figure}
 \includegraphics[width=0.43\textwidth]{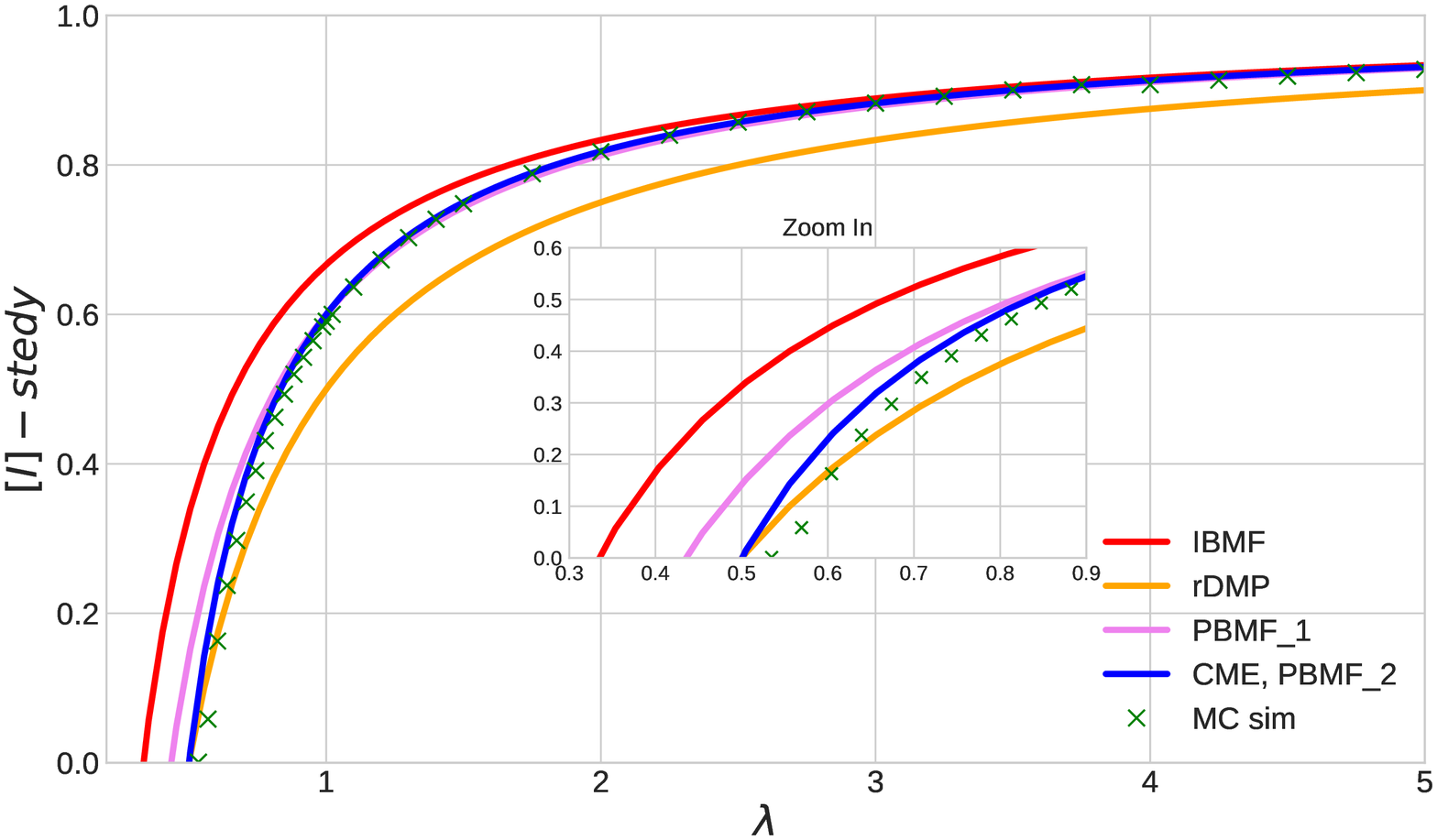}
  \includegraphics[width=0.43\textwidth]{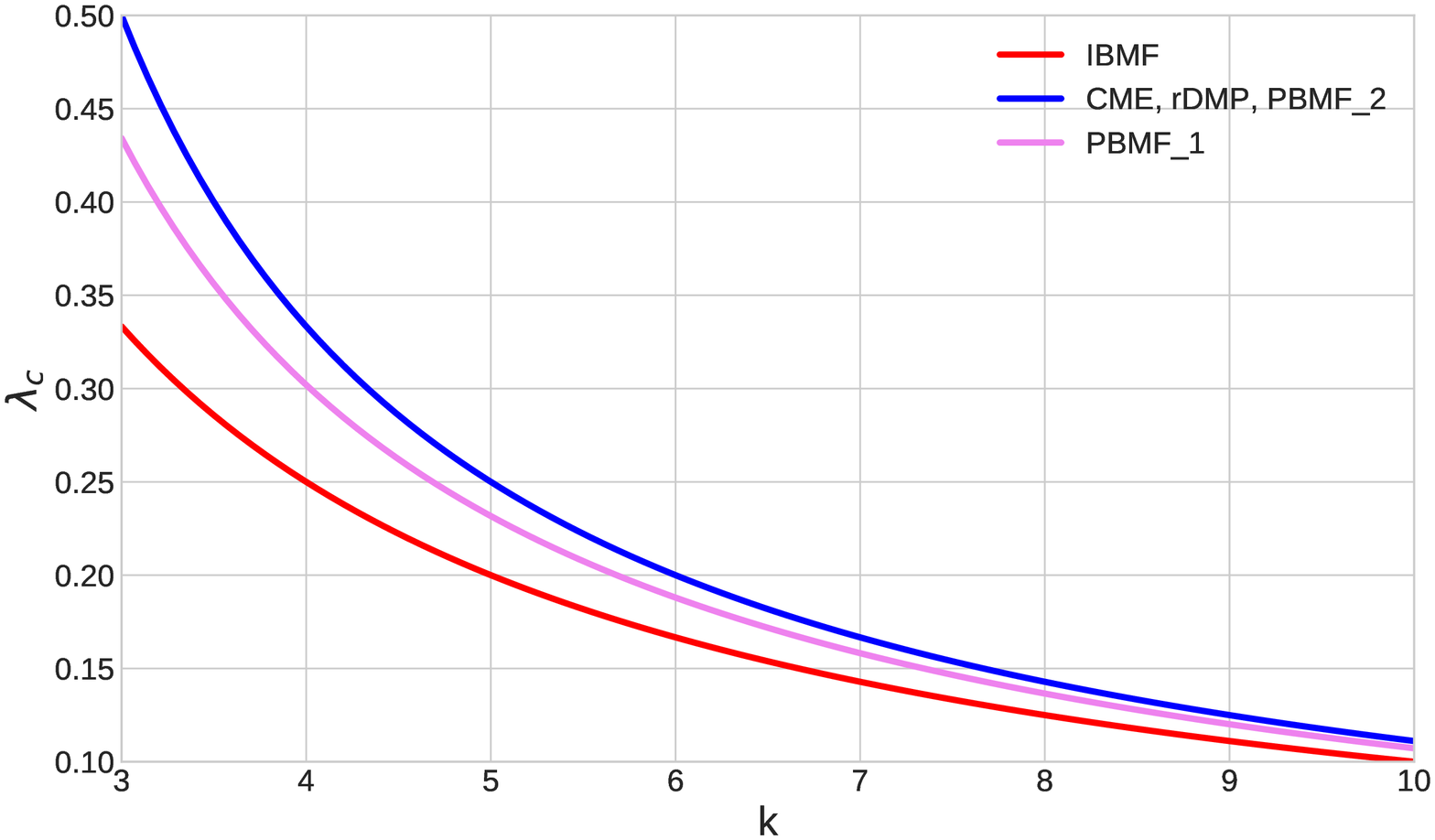}\caption{ {\bf Left}: Comparison between Monte Carlo simulations and the predicted endemic equilibrium for each approximation as a function of $\lambda$ for random regular graphs. Each MC point is an average of the last value of of the probability of infected computed for $10^4$ simulations over the 1000 nodes in the system. {\bf Rigth}: Comparison between the epidemic threshold predicted by the methods as a function of graphs connectivity. \label{fig:Eq_vs_R0}}
\end{figure}

In figure \ref{fig:Eq_vs_R0} we present the analytical predictions of each approximation for the epidemic size $p_i$ at the steady state as a function of $\lambda$. When compared to Monte Carlo results, mean field approximations (IBMF and PBMF-1) give an overestimation of the epidemic size at small $\lambda$, and rDMP an underestimation at large $\lambda$. Meanwhile, CME and PBMF-2 seems to be a better prediction mixing the large $\lambda$ behavior of IBMF and PBMF-1 with the small $\lambda$ behavior of rDMP.

The results presented in figure \ref{fig:Eq_vs_R0} are very similar to those shown in \cite{gleeson2013binary} where a tractable master equation is presented for the evolution in time of the epidemic size in a graph with the same degree. This description has more resemblance with a degree base aproximation and, as in our case, it catches with similar accuracy the value of the epidemic threshold for a random regular graph of conectivity 3.


%

\section{Average case}
\label{sec:average}
The systems of equations defining IBMF, PBMF, rDMP and CME could be large and delicate to solve on a given graph, although a simple numerical integration normally works. However, in many cases we are interested in general predictions for certain families of graphs or topologies. In this section we derive an average version of the CME approximation to characterize SIS epidemics on uncorrelated random graphs. Our approach relates closely to that in \cite{goltsev2012localization, van2012epidemic}, based on IBMF approximation, where the prevalence at the stationary state is derived just above the epidemic threshold. 

The simplest description of a graph ensemble is given by the distribution of degrees of its nodes. In the case of uncorrelated graphs, that distribution is the full description of the ensemble. One of the firts theoretical approaches used for epidemic modeling on  networks was the degree based mean field approach (DBMF) \cite{pastor2001epidemic}. It provided a set of master equations for the probability of a node of degree $k$ to be infected at time $t$, assuming statistical equivalence of all nodes of degree $k$. As stated in \cite{pastor2015epidemic} DBMF can be obtained by performing a degree-wise average over the IBMF equations. 

A similar procedure can be performed in the context of CME. Since the CME equations depend on the information coming from the neighbors in the network, it is expected that nodes more connected will have a different behavior than those less connected. We therefore attempt to reduce the number of equations in our system by characterizing all nodes with the same degree by a couple of average parameters 
\begin{equation}
 p^{\gamma}= \frac 1 {N^{\gamma}} \sum_{i:d_i=\gamma+1} p_i \qquad p_{\rightarrow}^{\gamma} = \frac 1 {M^\gamma} \sum_{i:d_i=\gamma+1} \sum_{j \in \partial i} p_{ij}^{\gamma} 
 \end{equation}
In both cases the normalization factors count the number of terms in the sums: $N^\gamma$ is the number of nodes with degree  $k = \gamma+1$ while $M^\gamma$ is the number of graph's edges that contain one of these $N^\gamma$ nodes. 

Averaging the equations (\ref{eq:CMESISpi}) and (\ref{eq:CMESISpij}), and after some simplifications, we get the average CME:
\begin{eqnarray}
\dot{p}^{\gamma}&=& -\mu p^{\gamma} + \beta (\gamma+1) (1-p^{\gamma}) \sum_{\gamma^{\prime}} g(\gamma^{\prime}) p_{\rightarrow}^{\gamma^{\prime}}\label{eq:average_pigamma} \\
\dot{p}_{\rightarrow}^{\gamma}&=& -\mu p_{\rightarrow}^{\gamma} + \beta \gamma (1-p_{\rightarrow}^{\gamma}) \sum_{\gamma^{\prime}} g (\gamma^{\prime}) p_{\rightarrow}^{\gamma^{\prime}} \label{eq:average_pijgamma}
\end{eqnarray}
where $g(\gamma)$ is a contact degree distribution. A node extracted randomly from the set of nodes $V$ has degree $k$ with distribution $k\sim P(k)$. However, in order to average the CME equations we rather need to know the excess-degree distribution $g(\gamma)$ of nodes that are sampled by randomly picking up an edge $(i,j)\in E$. Both distributions are related, assuming there is no further correlation in the graph, by:
 \begin{equation}
  g(\gamma)=\frac{(\gamma+1)P(\gamma+1)}{\sum_{\gamma}(\gamma+1)P(\gamma+1)} \quad \gamma\in[0,1\ldots].\label{eq:uncorrelated}
 \end{equation}

  \begin{figure}
 \includegraphics[width=0.7\textwidth]{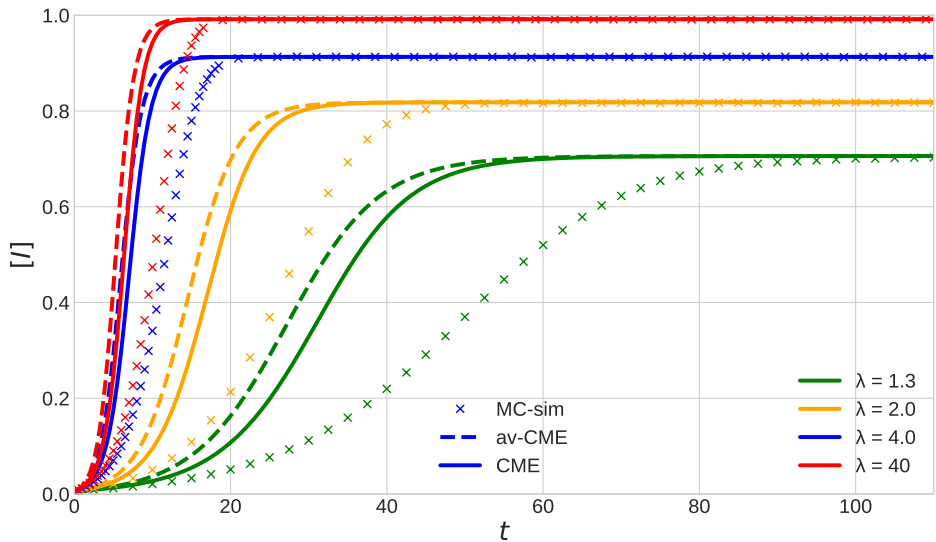}
  \caption{Average probablility of being infected as function of time. The epidemic outbreak was in 10 nodes of each graph of 1000 nodes in in an ensemble of 10 random regular graphs of connectivity 3. Each point of MC is an average over $10^4$ simulations.\label{fig:average}}
\end{figure}

Following \cite{pastor107066optimal, pastor2001epidemic} we can obtain the endemic threshold in terms of $\Theta = \sum_\gamma g(\gamma) p_{\rightarrow}^{\gamma}$. In the stationary state $\dot{p}^{\gamma} = 0$ and $\dot{p}_{\rightarrow}^{\gamma} = 0$ in (\ref{eq:average_pigamma}) and (\ref{eq:average_pijgamma}) leads to equations for $p^{\gamma}$ and  $p_{\rightarrow}^{\gamma}$ as a function of $\Theta$:
\begin{equation}
p^{\gamma}= \frac{\lambda (\gamma+1) \Theta}{1+\lambda (\gamma+1) \Theta } \qquad p_{\rightarrow}^{\gamma}= \frac{\lambda \gamma  \Theta}{1+\lambda \gamma \Theta } . \label{eq:pijtheta} 
\end{equation}
Plugging the last equation into the definition of $\Theta$, we obtain a self-consistency equation similar to the one in \cite{pastor107066optimal, pastor2001epidemic, pastor2001epidemic63}
\begin{equation}
 \Theta = f(\Theta) \equiv \sum_{\gamma} \frac{(\gamma+1)P(\gamma+1)}{\langle k \rangle } \frac{\lambda \gamma \Theta}{1+\lambda \gamma \Theta }  \label{eq:self-Theta}
\end{equation}
which always has a disease-free solution $\Theta = 0$. This solution becomes unstable (endemic case) when  $\partial_\Theta f(\Theta) |_0=1$, which defines the critical parameters resulting in
\begin{equation}
\lambda_c = \frac{\langle k\rangle}{\langle k^2\rangle- \langle k\rangle}    \label{eq:non-zero}
\end{equation}
where $\langle k \rangle $ is the average node degree. This result improves over the naive mean field prediction $\lambda)c = \langle k\rangle/\langle k^2\rangle$ \cite{pastor2015epidemic}. It trivially contains the result shown in table \ref{tb:table_endemic_state} for CME in random regular graphs and reduces to  it as $\langle k^2 \rangle = \langle k \rangle ^2$. For graphs with Poisson degree distribution (like Erdos-Renyi) $\langle k^2 \rangle = \langle k \rangle ^2 + \langle k \rangle$ the epidemic threshold becomes $ \lambda_c = \frac{1 }{\langle k\rangle}$.

\subsection{General graph ensembles}

Equations (\ref{eq:average_pigamma}) and (\ref{eq:average_pijgamma}) are already a reduction of $N$ differential equations to $K$ average equations, where $K$ is the maximum degree in the graph. However, $K$ itself could be large. We can further simplify by averaging now over the nodes degree and reducing to only two parameters $\tilde{p}_{\rightarrow} = \sum_{\gamma} g_{link}(\gamma) p_{\rightarrow}^{\gamma}$ and $\tilde{p} = \sum_{\gamma} P(\gamma) p^{\gamma}$:
\begin{eqnarray}
\dot{\tilde{p}}&=& -\mu \tilde{p} + \beta \tilde{p}_\rightarrow \sum_{\gamma} (\gamma+1) P(\gamma) (1-p^{\gamma}) \label{eq:me_av} \\
\dot{\tilde{p}}_{\rightarrow}&=& -\mu \tilde{p}_{\rightarrow} + \beta \tilde{p}_{\rightarrow} \sum_{\gamma} \gamma g_{link}(\gamma) (1-p^{\gamma}_{\rightarrow}) \label{eq:cme_av}
\end{eqnarray}
These equations are still not closed, since the right hand sides still depend on the degree based parameters. The product by $(\gamma+1)$ and $\gamma$ inside the sums in the right hand sides do not allow for a direct connection with the definitions of $\tilde p$ and $\tilde p_\rightarrow$. We will show, however, that in  the case of Erdos-Renyi graphs such connection can be obtained, though through some approximations and ansatzs. 
 
A very simple case is that of regular graphs, where degree distribution is deltaic $P(\gamma) = g(\gamma) =\delta_{k-1,\gamma}$. Equations (\ref{eq:average_pigamma}) and (\ref{eq:average_pijgamma}) reduce to two equations for the parameters $p^{k-1}(t)\equiv p(t)$ and $p^{k-1}_{\rightarrow}(t)\equiv p_{\rightarrow}(t)$:
 \begin{equation}
\dot{\tilde{p}}= -\mu \tilde{p} + \beta \:k \:\tilde{p}_{\rightarrow}  (1-p) \qquad \dot{\tilde{p}}_{\rightarrow}= -\mu \tilde{p}_{\rightarrow} + \beta \:(k-1) \:\tilde{p}_{\rightarrow}  (1-p_{\rightarrow})
\end{equation}
whose numerical integration can be compared with Monte Carlo simulations of epidemics in graphs with the same vertex degree $k$, and with the corresponding integration of the single instance CME equations (\ref{eq:CMESISpi}) and (\ref{eq:CMESISpij}). Figure \ref{fig:average} shows that the steady state is well predicted, while the transient is not, even in comparison with CME itself. This is a natural consequence of the lost of the spatial structure in the average case.

 
 
%
%
%

\subsection{Closure on Erdos Renyi graphs} \label{sub:closure_ER}

For an Erdos-Renyi graph node degrees are Poisson-distributed: $P(\gamma) = \frac{e^{- \kappa} \kappa^{\gamma + 1}}{(\gamma + 1) !}$ and $g_{link}(\gamma) = P(\gamma - 1)$, where $\kappa$ is the average degree.
We can connect the terms inside the sums in (\ref{eq:me_av}) and (\ref{eq:cme_av}) with the derivatives of $\tilde{p}$ and $\tilde{p}_{\rightarrow}$ with respect to the parameter $\kappa$ by noting that
\begin{eqnarray}
 \frac{\partial \tilde{p}}{\partial \kappa} 
&=   -\tilde{p} + \frac{1}{\kappa} \sum_{\gamma} (\gamma + 1) P(\gamma) \, p^{\gamma} \\
\frac{\partial \tilde{p}_{\rightarrow}}{\partial \kappa} 
 &=  -\tilde{p}_{\rightarrow} + \frac{1}{\kappa} \sum_{\gamma} \gamma \, g_{link}(\gamma) \, p_{\rightarrow}^{\gamma}
 \label{eqn:formal_p_der}
\end{eqnarray}
As it is shown in appendix \ref{ap:av_ER}, by substitution in (\ref{eq:me_av}) and (\ref{eq:cme_av}) we get:
\begin{eqnarray}
&& \dot{\tilde{p}} = - \mu \; \tilde{p} + \beta \, \kappa \, \tilde{p}_{\rightarrow} \left[1 - \tilde{p} - \frac{\partial \tilde{p}}{\partial \kappa} \right]  \label{eqn:ME_averaged_RGER} \\
&&\dot{\tilde{p}}_{_{\rightarrow}} = \left[ \beta \kappa - \mu \right] \tilde{p}_{\rightarrow} - \beta \; \kappa \; \tilde{p}_{\rightarrow} \left( \tilde{p}_{\rightarrow} + \frac{\partial \tilde{p}_{\rightarrow}}{\partial \kappa} \right)
 \label{eqn:CME_averaged_RGER}
\end{eqnarray}

\begin{figure}
\centering
\includegraphics[keepaspectratio=true,width=0.7\textwidth]{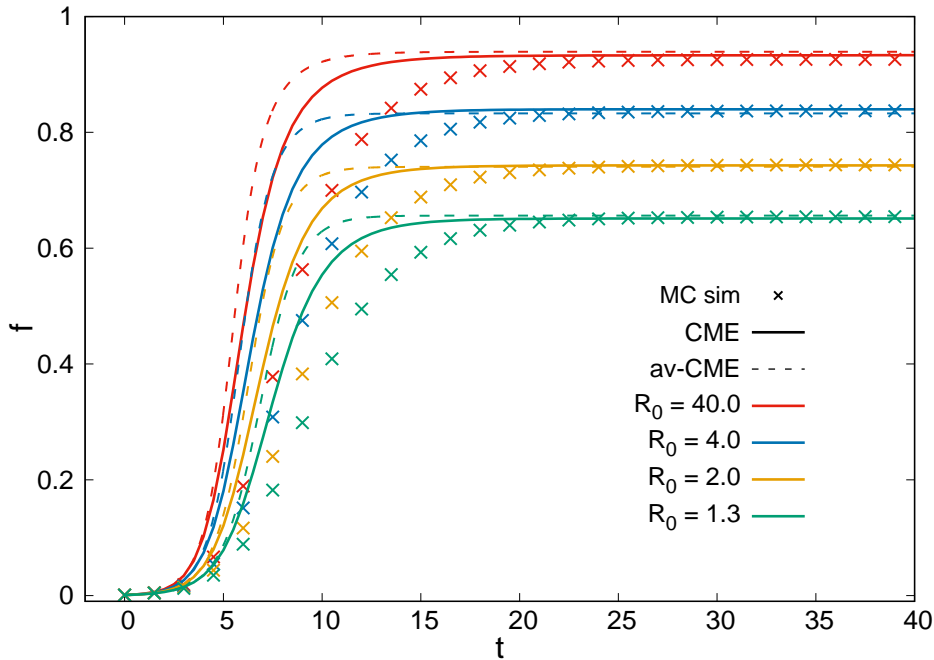}
\caption{ Comparison between average-CME and CME for Erdos Renyi graphs with mean connectivity (average degree) $\kappa=3$, several values of $\mu$ and $\beta = 0.4$. The figure shows the single site probability of infection as a function of time. As the ratio $\lambda=\beta / \mu$ decreases, the steady state has less infection probability.}
 \label{fig:av_CME_ansatz}
\end{figure}

In order to solve these equations we need some ansatz  for the dependence of the mean values  $\tilde{p}$ and $\tilde{p}_{\rightarrow}$ on the mean degree in the graph. Inspired by the whole derivation of the Cavity Master Equation for spin systems, we propose 
\begin{eqnarray}
\tilde{p}(t) = \frac{1}{2} \left[ 1 + \tanh{(\beta \, \kappa \, \chi(t) - \mu)} \right]  \label{eqn:ansatz_MF_single} \\
\tilde{p}_{\rightarrow}(t) = \frac{1}{2} \left[ 1 + \tanh{(\beta \, \kappa \, \epsilon(t) - \mu)} \right]
 \label{eqn:ansatz_MF_cav}
\end{eqnarray}
where $\chi(t)$ and $\epsilon(t)$ are some time-dependent fields that are obtained by inverting these very formulas in terms of $\tilde{p}$ and $\tilde{p}_{\rightarrow}$.

From this ansatz we can express the derivatives with respect to the degree in (\ref{eqn:ME_averaged_RGER}) and (\ref{eqn:CME_averaged_RGER}) in terms of $\tilde{p}$ and $\tilde{p}_{\rightarrow}$, respectively (see appendix \ref{ap:av_ER}). We get then a closed system of differential equations for these probabilities, that can be solved numerically. The results of the integration are shown in figure \ref{fig:av_CME_ansatz}.

\section{Extensions to SIR-SIRS models}
\label{sec:SIRSIRS}
Models with more compartments are also ubiquitous in epidemics modeling. In particular the SIR family, including SEIR and SIRS, that have experienced a boost in attention due to the COVID-19 actually ongoing epidemic. The Cavity Master Equation from \cite{aurell2017cavity} applies to spins $s_i=\pm 1$ physical models, and does not have an inmediate translation to general multi-states models. However, for the case of models with more than 2 states that occur in sequential order, as in 
\[S \rightarrow I \rightarrow R \rightarrow S\]
the steps in \cite{aurell2017cavity} are still valid, and CME equations are simple to derive.

The exact master equation for these sequential processes is still of the form (\ref{eq:originalME}) where the spin flip operator $F_i(\bsigma)$ produces a configuration where all variables keep their value, except for the $i$-th variable that is demoted to the preceeding state in the sequence. For instance, if $\sigma_i = R$, then $\sigma_i = I$ in $F_i(\bsigma)$.

The random point process representation  $(t_0,t_1,\ldots)$ of these multi-state stochastic models is still valid, and the transitions at time $t_i$ are unambigous given the sequential nature of the process. Following the steps in \cite{aurell2017cavity}, we get to the Cavity Master Equations that are a straight forward generalization of  (\ref{eq:CMEPi}) and (\ref{eq:CMEpij}) :
\begin{eqnarray}
\frac{\ud P_i (\sigma_i) }{\ud t} &=& \sum_{\sigma' \in\{\sigma_i,F(\sigma_i)\}} \sum_{\sigma_{\p i}} \Big[
(-1)^{\delta{\sigma_i,\sigma'}} r_i(\sigma', \sigma_{\p i}) \big[ \prod_{k \in \p i }
P_{ki}(\sigma_k| \sigma') \big] P_i(\sigma')
\Big]
\label{eq:CMEPiq} \\
 \frac{\ud p_{ij}(\s_i|\s_j)}{\ud t} &=&   \sum_{\sigma' \in \{\sigma_i,F(\sigma_i)\}} \sum_{\sigma_{\partial i\setminus j}} \Bigg[ 
(-1)^{\delta{\sigma_i,\sigma'}}  r_{i}[\sigma',\sigma_{\partial i}]  \Big[\prod_{k\in\partial i \setminus j } p_{ki}(\sigma_k|\sigma')\Big] p_{ij}(\sigma'|\s_j) 
 \Bigg]. \label{eq:CMEpijq}
\end{eqnarray}
Sums run over all q states of each variable ($q\in \{S,I,R\}$ for SIR and SIRS), and the $(-1)$ factor chooses accounts for whether we are arriving at state $\sigma_i$ or leaving it for the next sate. The functions $r_i(\sigma_i,\sigma_{\p i})$ at the rates at which the transition from state $\sigma_i$ occurs, depending also on the state of the neighbors variables.

For instance, in the SIR and SIRS cases the equations reduce to (after considering the explicit form of the rate functions):
\begin{eqnarray}
\frac{\ud P_i (\sigma_i\equiv I) }{\ud t} &=& -\mu P_i(I) + \beta P_i(S) \sum_{k\in \partial_i} p_{ki}(I | S) \label{eq:PiI}\\
\frac{\ud P_i (S) }{\ud t} &=& \gamma P_i(R) - \beta P_i(S) \sum_{k\in \partial_i} p_{ki}(I | S) \label{eq:PiS}\\
\frac{\ud P_i (R) }{\ud t} &=& -\gamma P_i(R) + \mu P_i(I) \label{eq:PiR}
\end{eqnarray}

\begin{eqnarray}
\frac{\ud p_{ij}(I|S) }{\ud t} &=& -\mu p_{ij}(I|S) + \beta p_{ij}(S | S) \sum_{k\in \partial_i\setminus j} p_{ki}(I | S) \label{eq:PijIS} \\
\frac{\ud p_{ij}(S|S) }{\ud t} &=& \gamma p_{ij}(R|S) - \beta p_{ij}(S|S) \sum_{k\in \partial_i\setminus j} p_{ki}(I | S) \\\frac{\ud p_{ij}(R|S) }{\ud t} &=& -\gamma p_{ij}(R|S) + \mu p_{ij}(I | S) 
\end{eqnarray}

Constants $\mu$ and $\beta$ are still the recovery rate and the infection rate, while $\gamma$ is the rate at which immunity is lost and patients pass from the recovered compartment back into the susceptible one. In  the case of SIRS model $\gamma>0$, while SIR corresponds to $\gamma=0$.

The first three equations (\ref{eq:PiI},\ref{eq:PiS},\ref{eq:PiR}) are still the same obtained by the dynamical message passing in \cite{shrestha2015message}
but, as in the SIS case, the conditional distribution equation (\ref{eq:PijIS}) differs from that in rDMP
\begin{eqnarray}
\frac{\ud p_{ij}(I|S) }{\ud t} &=& -\mu p_{ij}(I|S) + \beta P_j(S) \sum_{k\in \partial_i\setminus j} p_{ki}(I | S) \label{eq:rDMP_SIR_pij} 
\end{eqnarray}
in the factor $P_j(S)$ before the sum over neighbors. Notice that this difference implies that only four equations are needed in the rDMP case: 3 for each node $\dot{P_i(I)}, \dot{P_i(S)} , \dot{P_i(R)}$ and one on each directed edge $\dot{p_{ij}(I|S)}$; while in our case all 6 equations need to be integrated (3 on nodes, 3 on edges) since they are all mutually dependent.

\begin{figure}
 \includegraphics[width=0.48\textwidth]{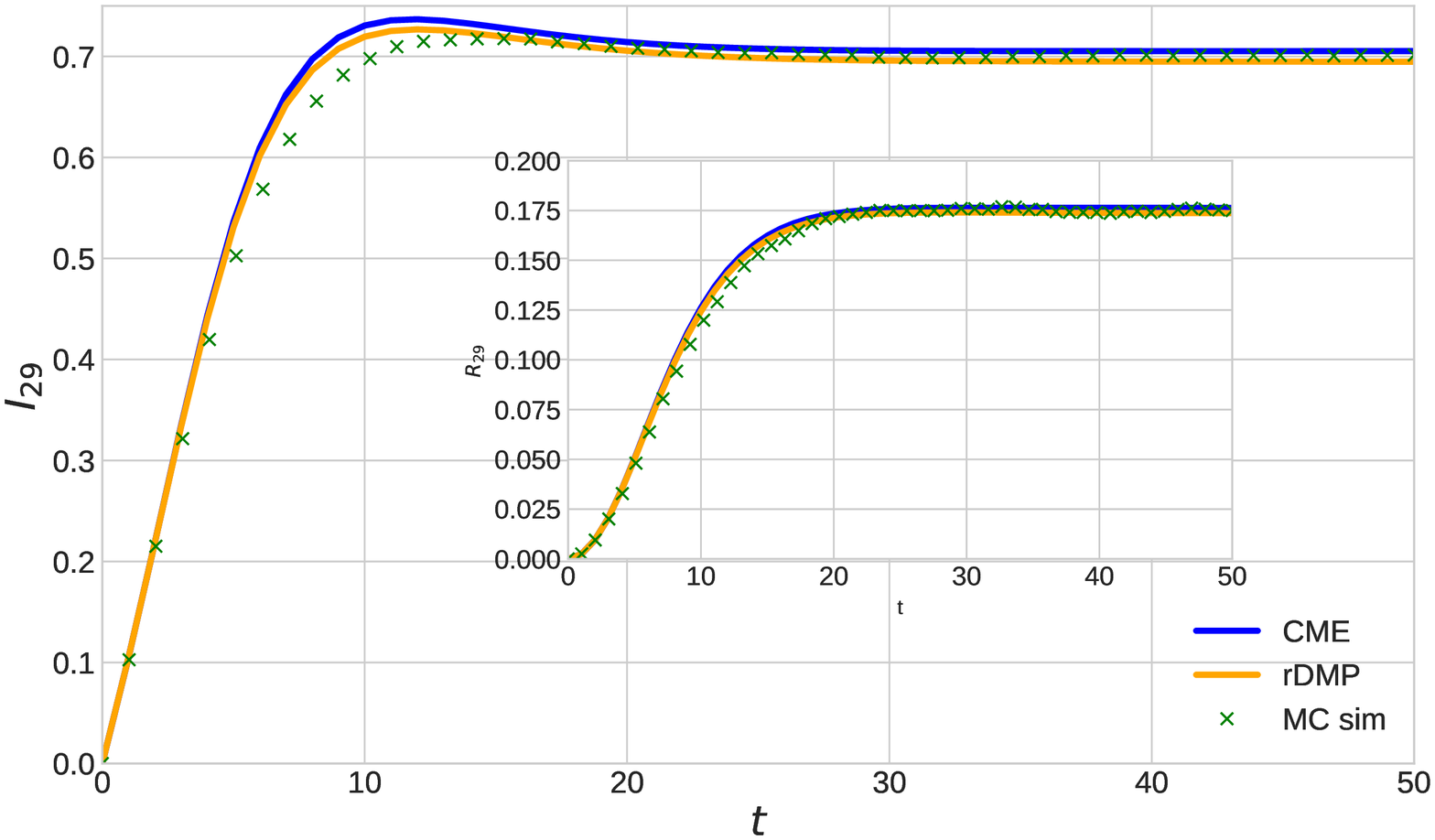}
 \includegraphics[width=0.48\textwidth]{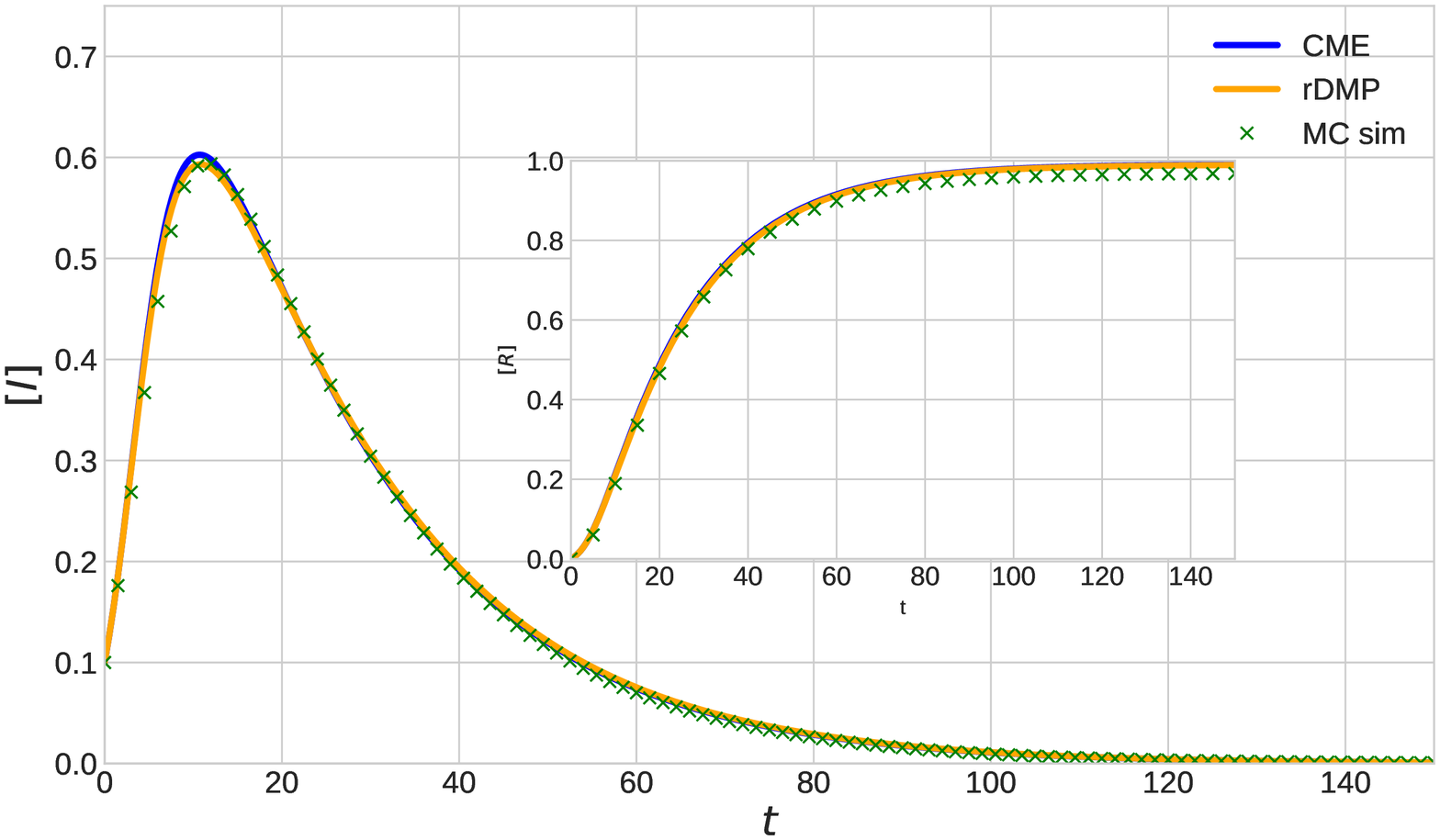}
 \caption{
{\bf Left}: Probability of node 29 of Zacharia's graph, with $\beta=0.1$, $\mu=0.05$ and $\gamma=0.2$, to be infected as function of time, discounting at each time the situations in which epidemic disappear in a SIRS model. The inerplot is the probability of being recovered. {\bf Right}: Percentage of the nodes in the infected state for a SIR model in a random graph of 1000 nodes with connectivity 5 and $\beta = 0.1$ and $\mu= 0.05$, starting with 100 infected persons . The inerplot shows the percentage of nodes in the recovered state. Each point of MC is an average over $10^5$ simulations.  \label{fig:zacSIRS}}
\end{figure}

Numerical experiments show that the CME approach fits better the average evolution of Monte Carlo simulations (figure \ref{fig:zacSIRS}) for SIRS and SIR models. 

\section{Conclusions}

The analytical and numerical results on susceptible-infectious-susceptible model have shown that cavity master equation (CME) is an effective approximation to the average dynamics of epidemic systems, where average is intended over many stochastic realizations of the epidemics process. 
CME corrects one term of the dynamic message passing equations, and this correction produces a more accurate prediction, when compared to numerical simulations. CME also outperforms an individual and a type of pair based mean field approximations. A different type of pair based closure appears to be equivalent to CME equations in the endemic (stationary) case, but better in the prediction of the transient.

We also explored with some success the average case (graph ensemble) predictions for random regular and Erdos-Renyi graphs, and briefly ended by extending CME to multi state models as SIR and SIRS.

Many aspects remain to be studied. We found particularly appeling the equivalence of CME and a pair based method in the steady state. Similarity is expected, since both approaches subsume first neighbors correlations by using the multivariate probability $p(s_i,s_j)$ or the conditional one $p(s_i|s_j)$. However, the fact that the pair based method is better at the transient, suggests  that there is room for improvement at CME. Understanding this could impact other applications of CME in the past and future. We also found surprising that CME is particulary bad in 1D ring networks. Exploring why and how to improve in such cases seems an appealing direction to us.

Finally, the extension of CME to Potts models with q states need to be formalized in the framework of point processes that was used to derive CME in the first place, of which the application to SIR case here is just a fortunate simple situation.

\section{Acknowledgements}

This research has received funding from the European Union’s Horizon 2020 research and innovation programme  MSCA-RISE-2016 under grant agreement No 734439 INFERNET and also from the PNCB ``Modelación matemática para la Epidemiología'' from the Cuban Science Ministry. We are thankful to Dr. Roberto Mulet and Dr. Alexei Vazquez, for useful discussions. 

\appendix

\section{Cavity master equation for SIS model.}\label{ap:average}

In this appendix we will present more details on the formulation of the CME for the SIS model. Let's take as starting points equations (\ref{eq:CMEPi}) and (\ref{eq:CMEpij}):
\begin{eqnarray}
\frac{\ud P (\sigma_i) }{\ud t} &=& - \sum_{\sigma_{\p i}} \Big[
r_i(\sigma_i, \sigma_{\p i}) \big[ \prod_{k \in \p i }
P(\sigma_k| \sigma_i) \big] P(\sigma_i)
-  r_i(-\sigma_i, \sigma_{\p i}) 
\big[ \prod_{k \in \p i }
P(\sigma_k| -\sigma_i) \big]P(-\sigma_i) \Big]
 \\
 \frac{\ud p(\s_i|\s_j)}{\ud t} &=& -  \sum_{\sigma_{\partial i\setminus j}} \Bigg[ r_{i}[\sigma_i,\sigma_{\partial i}]  \Big[\prod_{k\in\partial i \setminus j } p(\sigma_k|\s_i)\Big] p(\s_i|\s_j) 
-  r_{i}[-\sigma_i,\sigma_{\partial i}] \Big[\prod_{k\in\partial i \setminus j } p(\sigma_k|-\s_i)\Big] p(-\s_i|\s_j) \Bigg] 
\end{eqnarray}

Due to the complementarity of the probabilities $P_i(I)$ and $P_i(S)$ we can focus on just deriving the equation for one of the terms. Let's take $P_i(\sigma \equiv I)$ and rewrite equation (\ref{eq:CMEPi}) accordingly.
\begin{eqnarray}
\frac{\ud P_i (I) }{\ud t} &=& - \sum_{\sigma_{\p i}} \Big[
r_i(I, \sigma_{\p i}) \big[ \prod_{k \in \p i }
P_{ki}(\sigma | I) \big] P_i(I)
-  r_i(S, \sigma_{\p i}) 
\big[ \prod_{k \in \p i }
P_{ki}(\sigma_k| S) \big]P_i(S) \Big]
 \end{eqnarray}
The rate $r_i(I,\sigma_{\p i})$ is the transition rate from state $I \rightarrow S$ and in the SIS model and in most of epidemic models it does not depends on the state of the contacts of the Infected person. It just depends on the typical time of recovering (or die) from the disease, so this rate is directly the recovering rate $r_i(I,\sigma_{\p i})=\mu$. On the other hand $r_i(S,\sigma_{\p i})$ is the transition rate from $S \rightarrow I$. In infectious diseases, the contagion can only occur if a susceptible individual is in contact with an infected one ($r_i(S,\sigma_k \equiv I)= \beta$, $r_i(S,\sigma_k \equiv S)= 0$), and the probability of getting infected is additive with the number of infected contacts. This means that we can rewrite $r_i(S,\sigma_{\p i}) =\sum_{k \in \p i} r_i(S,\sigma_k)$.
\begin{eqnarray}
\frac{\ud P_i (I) }{\ud t} &=& -\mu P_i(I) \sum_{\sigma_{\p i}} 
   \prod_{k \in \p i }
P_{ki}(\sigma | I)  
+ P_i(S) \sum_{\sigma_{\p i}} \Big[  \sum_{k' \in \p i} r_i(S,\sigma_{k'}) 
\Big] \prod_{k \in \p i } 
P_{ki}(\sigma_k| S)
 \end{eqnarray}
Let's apply the equivalence $\sum_{\sigma_{\p i}}  \prod_{k \in \p i } \rightarrow  \prod_{k \in \p i }. \sum_{\sigma_{k}}$ to invert the sum and product in the first term of the right hand side (RHS).
\begin{eqnarray}
\frac{\ud P_i (I) }{\ud t} &=& -\mu  P_i(I) \prod_{k \in \p i } \Big[ \sum_{\sigma_{k}}  
P_{ki}(\sigma | I) \Big] +
P_i(S) \sum_{\sigma_{\p i}} \Big[  \sum_{k' \in \p i} r_i(S,\sigma_{k'}) 
\Big] \prod_{k \in \p i } 
P_{ki}(\sigma_k| S)
 \end{eqnarray}
The term $\sum_{\sigma_{k}} P_{ki}(\sigma | I)$ is exactly the sum of $P_{ki}(I | I) + P_{ki}(S | I) = 1$ and therefore also the product is equal to 1.
\begin{eqnarray}
\frac{\ud P_i (I) }{\ud t} &=& -\mu  P_i(I) +
P_i(S) \sum_{\sigma_{\p i}} \Big[  \sum_{k' \in \p i} r_i(S,\sigma_{k'}) 
\Big] \prod_{k \in \p i } 
P_{ki}(\sigma_k| S)
 \end{eqnarray}
Now we rewrite the second term of the RHS as follows:
\begin{eqnarray}
\frac{\ud P_i (I) }{\ud t} &=& -\mu  Pi(I) + P_i(S) \sum_{\sigma_{\p i}}  \sum_{k' \in \p i} r_i(S,\sigma_{k'}) P_{k'i}(\sigma| S) 
 \prod_{k \in \p i \setminus k'} P_{ki}(\sigma| S)  \\
\frac{\ud P_i (I) }{\ud t} &=& -\mu  Pi(I) + P_i(S) \sum_{\sigma_{k'}} \sum_{k' \in \p i} r_i(S,\sigma_{k'}) P_{k'i}(\sigma| S) 
 \sum_{\sigma_{\p i \setminus k'}}  \prod_{k \in \p i \setminus k'} P_{ki}(\sigma| S) 
 \end{eqnarray}
 Exchanging again the sum and the product, but now on the last term, we get $\sum_{\sigma_{\p i \setminus k'}} \Big[ \prod_{k \in \p i \setminus k'} P_{ki}(\sigma| S) \big] \Big] = 1$, and using that $r_i(S,S) = 0$ and $r_i(S,I) = \beta$ we obtain equation (\ref{eq:CMESISpi}):
 \begin{eqnarray}
  \frac{\ud p_i }{\ud t} &=&  - \mu p_i + \beta (1-p_i) \sum_{k} p_{ki}
    \end{eqnarray}
Following the same steps is easy to derive the equation (\ref{eq:CMESISpij}) for conditional probabilities.

\section{Average case equations for Erdos-Renyi graphs} \label{ap:av_ER}

In this appendix we will show how to perform the closure of average-case equations for Erdos-Renyi graphs (subsection \ref{sub:closure_ER}). Let's start by explicitly writing the derivatives with respect to $\kappa$:

\begin{eqnarray}
&& \frac{\partial \tilde{p}}{\partial \kappa} =  \frac{\partial }{\partial \kappa} \left( \sum_{\gamma} P(\gamma) p^{\gamma} \right)= \frac{\partial }{\partial \kappa} \left( \sum_{\gamma} \frac{e^{- \kappa} \kappa^{\gamma + 1}}{(\gamma + 1) !} p^{\gamma} \right)  \label{eqn:explicit_p_arrow_der} \\
&& \frac{\partial \tilde{p}_{\rightarrow}}{\partial \kappa} =  \frac{\partial }{\partial \kappa} \left( \sum_{\gamma} g_{link}(\gamma) p_{\rightarrow}^{\gamma} \right)= \frac{\partial }{\partial \kappa} \left( \sum_{\gamma} \frac{e^{- \kappa} \kappa^{\gamma}}{\gamma !} p_{\rightarrow}^{\gamma} \right)
 \label{eqn:explicit_p_der}
\end{eqnarray}

Computation of (\ref{eqn:explicit_p_arrow_der}) and (\ref{eqn:explicit_p_der}) gives:

\begin{eqnarray}
&& \frac{\partial \tilde{p}}{\partial \kappa} =  - \sum_{\gamma} \frac{e^{- \kappa} \kappa^{\gamma + 1}}{(\gamma + 1) !} p^{\gamma} +  \sum_{\gamma} (\gamma + 1) \frac{e^{- \kappa} \kappa^{\gamma}}{(\gamma + 1) !} p^{\gamma} = -\tilde{p} + \frac{1}{\kappa} \sum_{\gamma} (\gamma + 1) P(\gamma) p^{\gamma}  \label{eqn:computed_p_arrow_der} \\
&& \frac{\partial \tilde{p}_{\rightarrow}}{\partial \kappa} = - \sum_{\gamma} \frac{e^{- \kappa} \kappa^{\gamma}}{(\gamma) !} p_{\rightarrow}^{\gamma} +  \sum_{\gamma} \gamma \frac{e^{- \kappa} \kappa^{\gamma - 1}}{\gamma !} p_{\rightarrow}^{\gamma} = -\tilde{p}_{\rightarrow} + \frac{1}{\kappa} \sum_{\gamma} \gamma \, g_{link}(\gamma) p_{\rightarrow}^{\gamma}
 \label{eqn:computed_p_der}
\end{eqnarray}

The sums in the right hand sides of (\ref{eqn:computed_p_arrow_der}) and (\ref{eqn:computed_p_der}) are also involved in equations (\ref{eq:me_av}) and (\ref{eq:cme_av}). Then, remembering that $\kappa = \sum_{\gamma} (\gamma + 1) P(\gamma)$, we can rewrite equation (\ref{eq:me_av}) as follows:

\begin{eqnarray}
\dot{\tilde{p}}&=& -\mu \tilde{p} + \beta \tilde{p}_\rightarrow \sum_{\gamma} (\gamma+1) P(\gamma) (1-p^{\gamma}) \nonumber \\
\dot{\tilde{p}}&=&  -\mu \tilde{p} + \beta \tilde{p}_\rightarrow \sum_{\gamma} (\gamma+1) P(\gamma) - \beta \tilde{p}_\rightarrow \sum_{\gamma} (\gamma+1) P(\gamma) \,  p^{\gamma} \nonumber \\
\dot{\tilde{p}}&=&  -\mu \tilde{p} + \beta \kappa \, \tilde{p}_\rightarrow - \beta \kappa \tilde{p}_\rightarrow \left(\tilde{p} + \frac{\partial \tilde{p}}{\partial \kappa} \right) \label{eq:me_av_replaced}
\end{eqnarray}
which leads directly to equation (\ref{eqn:ME_averaged_RGER}). Equation (\ref{eqn:CME_averaged_RGER}) can be derived by an analogous procedure, using an equivalent expression for the mean connectivity: $\kappa = \sum_{\gamma} \gamma \, g_{link}(\gamma)$

Now we just need closed expressions for the derivatives in (\ref{eqn:ME_averaged_RGER}) and (\ref{eqn:CME_averaged_RGER}). In order to do so, let's compute the $\kappa$-derivative on both sides of ansatz (\ref{eqn:ansatz_MF_single}) and (\ref{eqn:ansatz_MF_cav}). We get:

\begin{eqnarray}
\frac{\partial \tilde{p}}{\partial \kappa} &=& \frac{1}{2}\left[ 1 - \tanh^{2}{(\beta \, \kappa \, \chi(t)  - \mu)} \right] \beta \, \chi(t)  \label{eqn:ansatz_MF_single_der} \\
\frac{\partial \tilde{p}_{\rightarrow}}{\partial k} &=& \frac{1}{2} \left[ 1 - \tanh^{2}{(\beta \, \kappa \, \epsilon(t) - \mu)} \right] \beta \epsilon(t)
 \label{eqn:ansatz_MF_cav_der}
\end{eqnarray}

We can re-use equations (\ref{eqn:ansatz_MF_single}) and (\ref{eqn:ansatz_MF_cav}) for eliminating $\chi(t)$ and $\epsilon(t)$ from (\ref{eqn:ansatz_MF_single_der}) and (\ref{eqn:ansatz_MF_cav_der}), thus obtaining the following closed expressions for the derivatives:

\begin{eqnarray}
\frac{\partial \tilde{p}}{\partial \kappa} &=& \frac{1}{2 \kappa}\left[ 1 - \left( 2 \tilde{p} - 1 \right)^{2} \right] \left[ \tanh^{-1} \left( 2 \tilde{p} - 1 \right) + \mu \right] \label{eqn:ansatz_MF_single_der_final} \\
\frac{\partial \tilde{p}_{\rightarrow}}{\partial k} &=& \frac{1}{2\kappa} \left[ 1 - \left( 2 \tilde{p}_{\rightarrow} - 1 \right)^{2} \right] \left[ \tanh^{-1} \left( 2 \tilde{p}_{\rightarrow} - 1 \right) + \mu \right]
 \label{eqn:ansatz_MF_cav_der_final}
\end{eqnarray}

This allows to numerically solve equations (\ref{eqn:ME_averaged_RGER}) and (\ref{eqn:CME_averaged_RGER}).

\bibliographystyle{unsrt}
\bibliography{biblio}

\end{document}